\documentclass[twocolumn,prl,aps]{revtex4}

\usepackage{epsfig}

\begin{document}
\textheight 237mm

\title{\Large Prediction of new crystal structure phases in metal
borides: a lithium monoboride analog to MgB$_2$.}

\author{Aleksey N. Kolmogorov and Stefano Curtarolo} 

\affiliation{ Department of Mechanical Engineering and Materials
Science, Duke University, Durham, NC 27708 }

\date{\today}

\begin{abstract}
Modern compound prediction methods can efficiently screen large
numbers of crystal structure phases and direct the experimental search
for new materials. One of the most challenging problems in alloy theory
is the identification of stable phases with a never seen prototype;
such predictions do not always follow rational strategies. While
performing {\it ab initio} data mining of intermetallic compounds we
made an unexpected discovery: even in such a well-studied class of
systems as metal borides there are previously unknown layered phases
comparable in energy to the existing ones. With {\it ab initio}
calculations we show that the new metal-sandwich (MS) lithium
monoboride phases are marginally stable under ambient conditions but
become favored over the known stoichiometric compounds under moderate
pressures. The MS lithium monoboride exhibits electronic features
similar to those in magnesium diboride and is expected to be a good
superconductor.
\end{abstract}

\maketitle


Development of theoretical methods able to guide the experimental
search for new materials with desired properties is a fast-growing
field of research in materials
science\cite{Ceder,Fontaine,Inverse,Evolution,Hart,SC1,Morgan,CALPHAD}.
A particular effort has been put into the coupling of {\it ab initio}
electronic structure methods with efficient data-mining algorithms to
determine and utilize correlations in binding mechanisms in a chosen
set of structures\cite{Evolution,Hart,SC1,Morgan}. However, even with
the most advanced optimization algorithms one has to restrict the
search space to a given structure\cite{Evolution}, lattice
type\cite{Hart}, or prototype library\cite{SC1,Morgan}. Moreover, in
some approaches structures are kept symmetry-constrained or not
relaxed at all. Considering the rich nature of bonding in solids, such
limitations can lead to overlooking the most stable phases. This makes
identification of new prototypes a vital step towards a more complete
description of alloys.

Expanding the library of {\it ab initio} energies of binary alloys
described in Ref. \cite{SC1} we observed that one of the phases in the
Mg-B system, A$_2$B$_2$ fcc-(111) (or V2 \cite{Zunger1}), unexpectedly
underwent significant structural relaxation and became comparable in
energy to the mixture of two coexisting stable phases, MgB$_2$ and
hcp-Mg. Upon examination of the relaxation process we found that
there is a continuous symmetry-conserving path from V2 (Fig. 1c) to a
new metal-sandwich (MS) structure MS1 (Fig. 1b)\cite{ab}. The latter
has four atoms per unit cell with hexagonal layers of boron separated
by two triangular layers of metal; it bears strong resemblance to the
AlB$_2$ prototype (Fig. 1a) having a similar $sp^2$ boron-boron
bonding. The extra metal layer is inserted in a close-packed fashion
with Mg-Mg bond length close to that in the pure hcp structure, which
may account for the near-stability (by a few meV/atom) of the
magnesium monoboride. To the best of our knowledge, this structure has
not been considered before for any binary alloy. We have constructed a
series of related MS structures with different stacking sequences of
metal and boron layers (such as MS2 in Fig. 1d\cite{abba}), which is a
generalization of the MS1 prototype.
\begin{figure}[b]
  \begin{center}
    \centerline{\epsfig{file=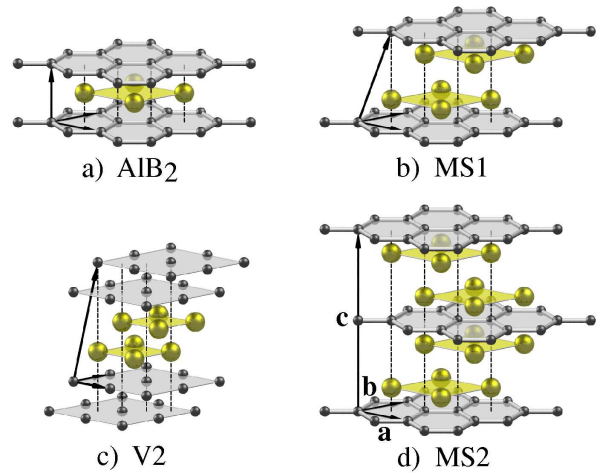,width=85mm,clip=}}
    \vspace{-1mm} \caption{\small (color online). Known (AlB$_2$,
    V2) and proposed metal-sandwich (MS1, MS2) prototypes. The
    hexagonal layers of boron (grey) are separated by triangular
    layers of metal (yellow, only four atoms per layer are
    shown).}
\label{mgb}
\end{center}
\end{figure}

The identification of the MS structures has prompted us to re-visit
metal borides, which have lately been a focal point in the search for
new superconducting materials, following the discovery of
superconductivity in MgB$_2$ with a surprisingly high transition
temperature of 39 K\cite{origin}. The high T$_c$ in MgB$_2$ was
explained as due to a very strong electron-phonon coupling of the
E$_{2g}$ mode to the hole-like $p\sigma$-states of
boron\cite{Kortus}. These states are also crucial for structural
stability of metal diborides and their availability in these compounds
implies weak binding\cite{Oguchi}. This feature makes the effectively
electron-doped (because of the extra metal layer) MS1 structure a
particularly good candidate to be a stable configuration for
low-valent metal borides. We have investigated a large set of alkali,
alkaline and transition metal borides with {\it ab initio}
calculations and found that MS1 and MS2 phases in one system, Li-B,
are indeed stable enough to compete against known phases.

Present {\it ab initio} simulations have been performed with {\small
VASP} \cite{kresse1993,kresse1996b}; we use projector augmented waves
(PAW)~\cite{bloechl994} and exchange-correlation functionals as
parametrized by Perdew, Burke, and Ernzerhof (PBE)\cite{PBE} for the
Generalized Gradient Approximation (GGA). For comparison purposes we
carry out tests in the Local Density Approximation (LDA)\cite{PZ}, and
with ultrasoft (US) pseudopotentials\cite{US}. We also employ an LAPW
code {\small WIEN2K} \cite{WIEN2K} to plot characters of electronic
bands and partial density of states (PDOS). All structures were fully
relaxed. Numerical convergence to within about 2 meV/atom is insured
by high energy cut-off (398 eV) and dense k-meshes\cite{k-points}.

Despite the potentially important application to batteries, the Li-B
system is not fully explored and the question of the most stable
monoboride phase remains open\cite{PF,aLi,bLi,B-Li}. Successful
experimental attempts to synthesize lithium monoboride under ambient
pressure date back to the 1970s\cite{Wan78,Wan79}. However, the
structure of $\alpha$-LiB, shown in Fig. 2a, has been identified only
recently\cite{B-Li,Wor00,aLi}. A theoretical study suggests that
lithium atoms in the lithium monoboride could be subject to low-energy
motion along the linear boron chains, as $\alpha$-LiB and a related
$\beta$-LiB phase (NiAs prototype, see Feg. 2b) were found to be
nearly degenerate\cite{bLi}.

We compare our MS lithium monoboride to the known stoichiometric
phases directly, using the experimentally observed $\alpha$-LiB as a
reference point. The following results may contain relatively big
systematic errors (i.e. due to approximations in the
exchange-correlation functional and in the treatment of the charge
density), since we consider structures with different types of
boron-boron bonding. We find that the MS1-LiB phase is 3 meV/atom
lower in energy than $\alpha$-LiB and about 7 meV/atom higher than
$\beta$-LiB\cite{CORE}. The $\alpha$-LiB and $\beta$-LiB phases differ by a
noticeable 10 meV/atom, a surprising result considering that our
fully-relaxed cell parameters ($\alpha$-LiB: $a=4.026$ \AA, $c=3.112$
\AA\ and $\beta$-LiB: $a=4.010$ \AA, $c=3.119$ \AA) closely match the
reported LDA values\cite{bLi}. We check the sensitivity of these
results to the approximation used by carrying out two more tests with
{\small WIEN2K} (LAPW, in the GGA-PBE) and {\small VASP} (PAW-LDA).

The relaxed cell parameters in the LAPW-GGA calculations are nearly
the same as in the PAW-GGA\cite{ab,abba}. In this approach MS1-LiB is
a little lower in energy (13 meV/atom below $\alpha$-LiB) and the most
stable of the three\cite{MT}. The PAW-LDA gives a much lower relative
energy for MS1-LiB (42 meV/atom below $\alpha$-LiB). A big discrepancy
in the description of relative energies with the two approximations
has also been seen in carbon systems with different types of
bonding\cite{carbon}; the GGA is expected to be more accurate. In all
the tests above $\alpha$-LiB and $\beta$-LiB remain 10 meV/atom
apart. The independence of this result on the calculation approach can
be explained by the similarities of the two structures, a trend also
observed in other systems\cite{al}.
\begin{figure}[t]
  \begin{center} \vspace{-5mm}
    \centerline{\epsfig{file=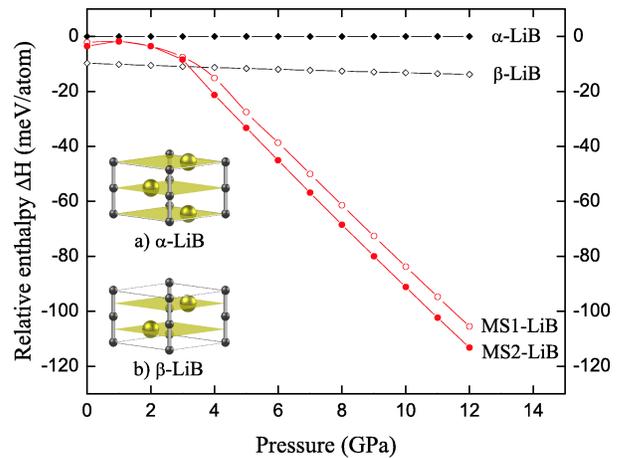,width=90mm,clip=}}
    \vspace{-8mm} \caption{ \small (color online). Calculated relative
    enthalpy, $\Delta$H(P)$\equiv$H(P)-H$_{\alpha-LiB}$(P), as a
    function of pressure for lithium monoboride phases. The points are
    connected with splines. The larger yellow spheres are lithium
    atoms and the smaller black spheres connected with sticks are
    boron atoms.}  \label{press} \end{center}
\end{figure}

We conclude that within the accuracy of these calculations the MS1-LiB
phase is comparable in energy to the known phases. Synthesis
conditions could therefore be the deciding factor for which compound
forms in the Li-B system at 50\% concentration. We next evaluate the
effect of hydrostatic pressure on the relative stability of the
lithium monoboride phases. The three phases have very close volumes
per atom at zero pressure. However, the $\alpha$-LiB and $\beta$-LiB
with linear boron chains are much harder than the $sp^2$-bound
MS1-LiB: the respective volumes at P = 12 GPa are about 85\%, 85\%,
and 69\% of the zero-pressure values (interestingly, the intralayer
bonds in MS1-LiB undergo non-monotonous expansion with pressure,
stretching by about 2\% at 5 GPa). The big compressibility of the
MS1-LiB is an expected feature for a layered structure with a weak
interlayer binding: at P = 12 GPa the c-axis shrinks down to 68\% of
its zero-pressure value, primarily due to the decrease in the Li-Li
interplanar distance. The weakening of the metallic binding in the
lithium bilayer is apparently caused by the significant charge
transfer from lithium to the boron layer. The extreme softness of
MS1-LiB suggests use of high pressures for synthesis of this phase. We
plot the calculated enthalpies of the considered lithium monoboride
phases with respect to $\alpha$-LiB versus external pressure in
Fig. \ref{press}. The plot shows that even if the MS-LiB are not the
most stable phase at ambient pressure, they are quickly driven below
the known phases and become favored by over 90 meV/atom at P = 12
GPa\cite{PRESS}. Note, that the hydrostatic pressure removes the
degeneracy of the MS1-LiB and MS2-LiB, different only by a long-period
stacking shift: the pressure favors MS2-LiB, in which the flatness of
the boron layers is enforced by symmetry. Use of appropriate
substrates could also promote growth of the new layered structures.
\begin{figure}[t]
  \begin{center} \vspace{-5mm}
    \centerline{\epsfig{file=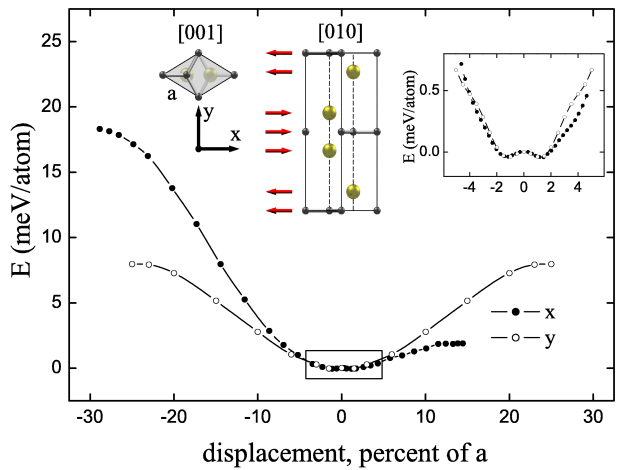,width=90mm,clip=}}
    \vspace{-8mm} \caption{ \small (color online). Slices of the
    potential energy surface in MS2-LiB along the eigenvectors of the
    softest phonon modes at $\Gamma$. The modes represent the
    interlayer sliding in the x (see the [010] view) and y directions
    of undistorted boron layers rigidly locked with the adjacent
    lithium layers. The inset demonstrates two shallow minima at small
    displacements. {\it Ab initio} data points are connected with
    splines.}  \label{soft} \end{center}
\end{figure}

An important question is how to detect the new phases if they indeed
form. Because the MS1 and MS2 are nearly degenerate at low pressures,
the resulting structure at 50\% concentration in Li-B could be a
random mixture of the two and would be hard to detect with standard
x-ray methods. Fortunately, the two phases have similar interlayer
distances for boron and one would expect to see an x-ray peak (at
$\lambda=1.5418$ \AA, zero pressure) in the range from
$2\theta=16.1^\circ$ for MS2 to $2\theta=16.6^\circ$ for MS1. None of
the observed peaks in the samples prepared at ambient
pressure\cite{aLi} match these calculated values. Experimental effort
to synthesize lithium monoboride at high pressures is highly desired.

{\it Ab initio} studies have shown that existence of energetically
favorable phases could be barred by dynamical
instabilities\cite{UNSTABLE}. We have investigated phonon modes at
$\Gamma$-point with a force-constant matrix diagonalization method,
implemented in {\small VASP}, using the MS2 structure with a
convenient rhombohedral unit cell. MS2-LiB has two weakly interacting
boron layers per unit cell which leads to an effective
double-degeneracy of the known phonon modes at $\Gamma$ in the
AlB$_2$-type compounds. Three of the remaining six optical modes,
which involve sliding of undistorted layers, are rather soft. The soft
modes are defined by the weakened interlayer Li-Li interaction and the
perfectly stacked structure may actually be prone to small interlayer
shifts, as there are two shallow minima at about 2\% displacements of
the in-plane lattice vector a (see Fig. 3). We crudely estimate the
frequency of the soft modes at $\Gamma$ by using all points up to 5\%
displacements and obtain $\omega_{x,y}$=55 cm$^{-1}$ and $\omega_z$=86
cm$^{-1}$. The modes at the A point are slightly softer
($\omega_{x,y}$=37 cm$^{-1}$ and $\omega_z$=59 cm$^{-1}$) but they
remain real (interestingly, the frequencies are comparable to those in
another layered structure, graphite\cite{Wakabayashi}). The interlayer
sliding phonon modes along the $\Gamma$-A direction are expected to be
the softest optical phonon modes in the whole Brillouin zone: any
finite $k_{x,y}$ would result in distortions of the rigid boron or
lithium layers and a consequent phonon hardening. These considerations
indicate that the MS2-LiB phase is dynamically stable. Hydrostatic
pressure is again found to be a stabilizing factor: already at P=2 GPa
the small distortions are suppressed and the frequencies of the
softest x-y phonon modes are nearly doubled. Calculation of phonon
modes for the full Brillouin zone is a subject of future study. With
all the phonon branches calculated one can also estimate the
vibrational entropy contribution to the Gibbs free energy to complete
the analysis of the thermodynamic stability of MS-LiB.

The investigation of the phonon modes reveals that the actual stable
unit of the MS-LiB phases is a hexagonal boron layer coated with two
lithium layers on each side and can be related to graphene. Indeed,
according to our estimates the cohesion per area between these
B$_2$Li$_2$ sheets and their curvature energy are comparable to those
for a carbon layer\cite{al}. One could speculate that if MS-LiB was
exfoliated into such sheets, they might form carbon-like tubular or
nanoporous structures. It would be interesting to see whether stable
MS compounds could be obtained in other intermetallics based, for
instance, on beryllium. Going beyond binaries also seems a promising
direction to find more stable compounds: additional metal sites give
more flexibility to experiment with different ternary alloys.

Finally, we give a brief discussion of the electronic properties of
the MS phases focusing primarily on the $p$-states in boron, important
for superconductivity in metal diborides. Our LAPW calculations
\cite{WIEN2K,lapw} show that the boron $p\sigma$ band along $\Gamma$-A
in MS2-LiB has practically no dispersion (Fig 3), an expected feature
caused by the large separation between boron layers. Compared to a
hypothetical AlB$_2$-LiB$_2$ phase the $\Gamma$-A band downshifts by
0.6 eV and the $p\sigma$ PDOS at Fermi level drops by 25\% to
$N_{p_{xy}}^B(0)=0.046$ states/(eV$\cdot$spin) per boron atom, which
indicates higher filling of these bonding boron states. Note, that
bonding $p\pi$ states also become occupied (E$_F$ is located exactly
in the middle of the $p\pi$ pseudo-gap). The combination of these
effects explains the dramatic stabilization of the lithium
monoboride. Remarkably, the $p\sigma$ PDOS in MS2-LiB stays about 12\%
higher than that in MgB$_2$.
\begin{figure}[t]
  \begin{center} \vspace{-5mm}
    \centerline{\epsfig{file=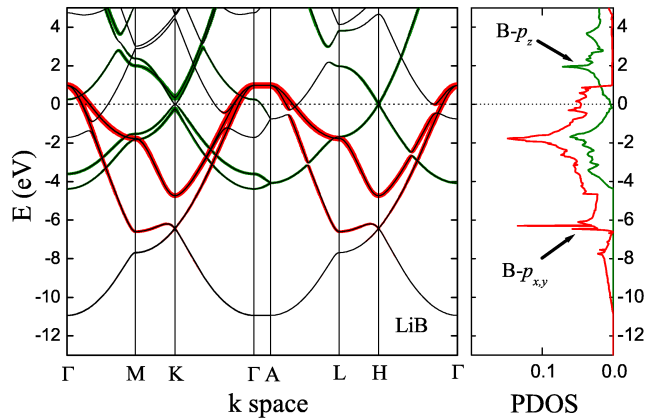,width=95mm,clip=}}
    \vspace{-8mm} \caption{ \small (color online). Band structure and
    partial density of states (PDOS) in MS2-LiB, calculated in
    LAPW\cite{WIEN2K,lapw}. PDOS units are states/(eV$\cdot$spin) per
    boron
    atom. The thickness of band structure lines is proportional to
    boron $p_{x,y}$ (red) and $p_z$ (green) characters.}
    \label{bsdos} \end{center}
\end{figure}

We find the frequency of the anharmonic E$_{2g}$ mode at
$\Gamma$-point (649 cm$^{-1}$) in MS2-LiB to be quite close to that
(624 cm$^{-1}$) in MgB$_2$\cite{Roxana}, which can be related in part
to the almost perfect match of the covalent boron-boron bond
lengths\cite{abba}. The similarities of the boron layer properties in
these borides appear to extend to the electron-phonon coupling as
well: following procedure in Ref. \cite{EPH}, we estimate the
deformation potential $\mathcal{D}$ for the E$_{2g}$ mode to be 13
eV/\AA\ in both phases. Unlike the two-band superconductor MgB$_2$,
MS2-LiB has negligible contribution from the boron $p\pi$ states at
E$_F$. These simple arguments suggest that the superconducting T$_c$
in the MS2-LiB would be at least comparable to that in MgB$_2$.

In conclusion, this work demonstrates that our knowledge of binary
alloys is still incomplete and the identification of yet unknown
phases may require new prediction strategies. The rather accidental
finding of the MS phases should be credited to the exhaustive
consideration of all candidates used in our data-mining approach and
the careful structural relaxation in the calculation of their ground
states. The MS structures are shown to be particularly suitable for
the Li-B system: lithium monoborides are marginally stable under
ambient conditions but become favored over the known stoichiometric
compounds under pressure. The effectively electron-doped MS-LiB phases
retain a significant PDOS from the $p\sigma$-boron states at E$_F$ and
could be the long-sought non-magnesium-based high T$_c$ metal
boride. Moreover, relative to MgB$_2$ the MS-LiB phases are
demonstrated to be hole-doped; if synthesized, they would provide
valuable information on how the hole-doping affects the T$_c$ in
stoichiometric metal borides and whether MgB$_2$ really has the
highest possible T$_c$ in its
class\cite{LiBC,LixBC,dope_review,peihong}.

We thank F.H. Cocks, V. Crespi, P. Lammert, R. Margine, and J. Sofo
for valuable discussions.


\begin{thebibliography}{99}

\bibitem{Ceder}
G. Ceder {\it et al.}, Nature (London) {\bf 392}, 694 (1998).

\bibitem{Fontaine}
D. de Fontaine, in {\it Solid State Physics}, edited by H. Ehrenreich
and D. Turnbull (Academic Press 1994), {\bf 47}, pp. 33-176.

\bibitem{Inverse}
A. Franceschetti and A. Zunger, Nature {\bf 402}, 60 (1999).

\bibitem{Evolution}
G.H. Johannesson {\it et al.}, Phys. Rev. Lett. {\bf 88}, 255506
(2002).

\bibitem{Hart}
G.L.W. Hart {\it et al.}, Nature Materials {\bf 4}, 391 (2005).

\bibitem{SC1}
S. Curtarolo {\it et al.}, Phys. Rev. Lett. {\bf 91}, 135503 (2003).

\bibitem{Morgan}
D. Morgan {\it et al.}, Meas. Sci. Technol. {\bf 16}, 296 (2005).

\bibitem{CALPHAD}
S. Curtarolo {\it et al.}, Calphad 29, 163-211 (2005).

\bibitem{Zunger1}
R. Magri {\it et al.}, Phys. Rev. B {\bf 42}, 11388 (1990).

\bibitem{ab}
MS1: 4 atoms in the primitive unit cell, space group R$\bar{3}$m
(\#166). $a=b$, $\alpha=90^\circ$, $\beta=90^\circ$,
$\gamma=120^\circ$, Wyckoff positions: M (0,0,$1/6-z_M/6$), B
(0,0,1/3+$\delta$). LiB: a=$3.058$ \AA, c$=16.06$ \AA, $z_M=0.485$,
$|\delta|<10^{-3}$.

\bibitem{abba} 
MS2: 8 atoms, space group P63/$mmc$ (\#194).  $a=b$,
$\alpha=90^\circ$, $\beta=90^\circ$, $\gamma=120^\circ$, Wyckoff
positions: M ($4f$) (1/3,2/3,$1/4-z_M/4$), B1 ($2b$) (0,0,1/4), B2
($2d$) (1/3,2/3,3/4). a$= 3.057$ \AA, c$ = 11.04$ \AA, $z_M=0.496$.

\bibitem{origin}
J. Nagamatsu {\it et al.}, Nature {\bf 410}, 63 (2001).

\bibitem{Kortus}
J. Kortus {\it et al.}, Phys. Rev. Lett. {\bf 86}, 4656 (2001).

\bibitem{Oguchi}
T. Oguchi, J. Phys. Soc. Jpn. {\bf 71}, 1495 (2002).

\bibitem{kresse1993}
G. Kresse and J. Hafner, Phys. Rev. B {\bf 47}, 558 (1993).

\bibitem{kresse1996b}
G. Kresse and J. Furthmuller, Phys. Rev. B {\bf 54}, 11169
(1996).

\bibitem{bloechl994}
P. E. Blochl, Phys. Rev. B {\bf 50}, 17953 (1994).

\bibitem{PBE}
J.P. Perdew {\it et al.}, Phys. Rev. Lett. {\bf 77} 3865 (1996).

\bibitem{PZ}
D.M. Ceperley and B.J. Alder, Phys. Rev. Lett. {\bf 45}, 566 (1980);
J.P. Perdew and A. Zunger, Phys. Rev. B {\bf 23}, 5048 (1981).

\bibitem{US}
D. Vanderbilt, Phys. Rev. B {\bf 41}, R7892 (1990).

\bibitem{WIEN2K}
An augmented plane wave+local orbitals program for calculating crystal
properties P. Blaha, K. Schwarz, G.K.H. Madsen, D. Kvasnicka, and
J. Luitz, {\small WIEN2K}, Karlheinz Schwarz, Technical Universitdt
Wien, Austria, 2001.

\bibitem{k-points} We use 20$\times$20$\times$23 for $\alpha$-LiB and
$\beta$-LiB and 24$\times$24$\times$12 for MS1-LiB (with
$\Gamma$-point included) meshes\cite{MONKHORST_PACK}. Quadrupling the
total number of k-points changes the relative energies by less than 1
meV/atom.

\bibitem{MONKHORST_PACK}
J.D. Pack and H.J. Monkhorst, {Phys. Rev. B} {\bf 13}, 5188 (1976);
{\bf 16}, 1748 (1977).

\bibitem{PF}
P. Villars, K. Cenzual, J. L. C. Daams, F. Hulliger,
T. B. Massalski, H. Okamoto, K. Osaki, A. Prince, and S. Iwata,
Crystal Impact, {\it Pauling File. Inorganic Materials Database and
Design System}, Binaries Edition, ASM International, Metal Park, OH
(2003).

\bibitem{aLi}
Z. Liu {\it et al.}, J. Alloys Compd. {\bf 311}, 256 (2000).

\bibitem{bLi}
H. Rosner and W.E. Pickett, Phys. Rev. B {\bf 67}, 054104 (2003).

\bibitem{B-Li}
H.B. Borgstedt and C. Guminski, J. Phase Equilibria {\bf 24} 572 (2003).

\bibitem{Wan78}
F.E. Wang {\it et al.}, J. Less-Common Met. {\bf 61}, 237 (1978).

\bibitem{Wan79}
F.E. Wang, Metall. Trans. A {\bf 10}, 343 (1979).

\bibitem{Wor00}
M. Worle and R. Nesper, Angew. Chem. Int. Ed. {\bf 39}, 2349 (2000).

\bibitem{CORE}
Due to the ionization of lithium atoms in Li-B compounds the 1s
semi-core states in Li are likely to relax. Thus, in all PAW
calculations we use the potentials where the semi-core 1s states
treated as valence. Note, that with regular PAW potentials (1s states
are considered as core) the MS1-LiB is 12 meV/atom below
$\alpha$-LiB.

\bibitem{MT} 
We use $R_{MT}^{B}=1.35$ a.u., $R_{MT}^{Li}=1.8$ a.u., $R_{MT}\cdot
K_{max}=8.0$, $l_{max}=10$ and require the charge convergence to
within 10$^{-4}$. Since boron-boron bond lengths in the linear chains
and in the hexagonal layers are quite different (by about 13\%) we
also check the dependence of the results on the muffin-tin radii
used. Increasing $R_{MT}^{B}$ to $1.45$ a.u. in $\alpha$-LiB,
$R_{MT}^{B}$ to $1.6$ a.u. in MS1-LiB and $R_{MT}^{Li}$ to $2.0$
a.u. in both (the MT spheres do not overlap) changes the relative
energies by less than 2 meV/atom. LAPW and APW+lo\cite{WIEN2K} basis
sets produce nearly identical results.

\bibitem{carbon}
A. Catellani {\it et al.}, Phys. Rev. B {\bf 62}, R4794 (2000).

\bibitem{al}
A.N. Kolmogorov and V.H. Crespi, Phys. Rev. B {\bf 71} 235415 (2005).

\bibitem{PRESS} The formation enthalpy with respect to
$\alpha$-B\cite{Oguchi} and fcc-Li becomes even more negative in the
considered range of pressures. The known boron-rich
compounds LiB$_3$ and Li$_3$B$_{14}$\cite{B-Li,PF} have large unit
cells with fractional occupancies and cannot be presently evaluated
with desired degree of accuracy.

\bibitem{UNSTABLE}
A. Zunger {\it et al.}, Phys. Stat. Sol. B {\bf 223}, 369-378 (2001).

\bibitem{Wakabayashi}
R. Nicklow {\it et al.}, Phys. Rev. B {\bf 5}, 4951 (1972).

\bibitem{lapw}
$R_{MT}^{Li}=2.0$ a.u., $R_{MT}^{B}=1.6$ a.u., and $R_{MT}\cdot
K_{max}=8.0$.

\bibitem{Roxana} 
We reproduce the 601 cm$^{-1}$ anharmonic frequency for MgB$_2$
obtained in US pseudopotential calculations\cite{PHONONS}. With the
same simulation settings but in the PAW approach the frequencies are
23 cm$^{-1}$ higher. We expect the {\it difference} of the frequencies
in MgB$_2$ and MS-LiB to be more accurate than the absolute values,
which might not be properly evaluated in the frozen-phonon
approach\cite{ANHARM}.

\bibitem{PHONONS}
K. Kunc {\it et al.}, J. Phys.: Condens. Matt. {\bf 13}, 9945-9962 (2001).

\bibitem{ANHARM}
M. Lazzeri {\it et al.}, Phys. Rev. B {\bf 68}, 220509 (2003).

\bibitem{EPH}
J.M. An and W.E. Pickett, Phys. Rev. Lett. {\bf 86}, 4366 (2001).

\bibitem{LiBC}
H. Rosner {\it et al.}, Phys. Rev. Lett. {\bf 88}, 127001 (2002).

\bibitem{LixBC}
A.M. Fogg {\it et al.}, Chemical Communications {\bf 12} 1348 (2003).

\bibitem{dope_review}
R.J. Cava {\it et al.}, Physica C {\bf 385}, 8 (2003).

\bibitem{peihong} 
P. Zhang {\it et al.}, Phys. Rev. Lett. {\bf 94}, 225502 (2005).

\end{thebibliography}
\end{document}